\documentclass[reprint,aps,showkeys,nofootinbib]{revtex4-2}

\usepackage{graphicx}
\usepackage{dcolumn}
\usepackage{bm}
\usepackage{hyperref}
\usepackage{amsmath}
\usepackage{amssymb}
\usepackage{amsfonts}
\usepackage{resizegather}
\usepackage{caption}
\usepackage[british]{babel}
\usepackage{physics}

\usepackage{color}
\usepackage{ulem}
\usepackage{cancel}
\usepackage[export]{adjustbox}
\usepackage[newitem,newenum,defblank]{paralist}

\newcommand{\lmatt}{\mathcal{L}_\text{matt}}

\begin{document}

\title{The matter Lagrangian of a non-perfect fluid}

\author{Sergio Mendoza}
\email{sergio@astro.unam.mx}
\author{Sarah\'{\i} Silva}
\email{sgarcia@astro.unam.mx}
\affiliation{Universidad Nacional Aut\'onoma de M\'exico, Instituto de 
Astronom\'{\i}a, Ciudad Universitaria AP 70-264, Ciudad de M\'exico 04510, 
M\'exico}

\date{\today}


\begin{abstract}
  We show that the matter Lagrangian of a non-perfect fluid takes the
negative or positive value of the total energy density of the fluid, which
is composed of the the rest plus internal energy densities. The ambiguity
of the sign depends on the definition of the matter action and the chosen
signature.
\end{abstract}


\maketitle

\section{Introduction}

  Since the mid 1950's it has become important to compute the value of
the matter Lagrangian of an ideal fluid~\citep[see e.g.][and references
therein]{mendoza2021matter}.  Over the years, different approaches have
shown that this value can be either (up to a sign) the pressure or its
full energy density, this last one composed of the addition between the
rest energy density plus the thermodynamical internal energy density of
the fluid.  \citet{mendoza2021matter} showed that the only possible value
that the matter Lagrangian of an ideal fluid can only be the complete
energy density of the fluid (up to a sign depending on definitions of
the matter action and the signature of the metric).

  Using calculus of variations, in this article we show that the
matter Lagrangian $\lmatt$ of a non-perfect fluid depends on the total
energy density of the fluid $e$ in the form $\lmatt=\mp e$, where the sign
of the matter Lagrangian depends on the definition of the matter action
and on the signature of the metric. In other words, the matter Lagrangian
of a non-perfect fluid consists of the rest mass energy density plus the
purely thermodynamical internal energy density of the fluid.  
In Section~\ref{action-section} we
present the matter action and the definition of the energy-momentum through
the variations of this action.  In Section~\ref{Fluids} we introduce some
basic thermodynamic relations and the energy momentum tensor of a non
perfect fluid in the Eckart frame. In Section~\ref{matter-lagrangian}
we use the mass continuity equation and the energy equation in the
Eckart frame in order to find the value of the matter Lagrangian of a 
non-perfect fluid.  Finally in Section~\ref{remarks} we discuss our results
and show that the same result for the matter Lagrangian case is also valid
in the Landau frame of the fluid.

\section{Action}
\label{action-section}

The matter action \( S_\text{matt} \) is defined as~\citep[see
e.g.][]{landau-fields}:

\begin{equation}
  S_\text{matt} = \pm \frac{ 1 }{ c } \int{ \lmatt \, \sqrt{-g} \, \mathrm{d}^4 x },
\label{action}
\end{equation}

\noindent where $ \lmatt $ represents the matter Lagrangian, $c$
is the speed of light an $g$ is the non-positive determinant of the
metric tensor $g_{\alpha \beta}$.  The choice of a $+$ or $-$ sign in 
equation~\eqref{action} is a definition convention. In this work we use a
signature (\(+,-,-,-\)) for the metric.  Greek space-time indices vary
from \( 0 \) to \( 4 \) and spatial Latin indices vary from \( 1 \)
to \( 3 \) and we use the  Einstein's summation convention over repeated
indices.

  The metric, Hilbert or Belifante-Rosenfeld energy-momentum tensor $ T_{\alpha\beta} $ is defined through the variations $ \delta S $ of the matter 
action~\eqref{action}. Assuming that the matter
Lagrangian is a function of the metric tensor $ g^{\alpha\beta}$ only 
and not of its
derivatives $ \partial g^{\alpha\beta} / \partial x^\lambda $, the metric energy-momentum tensor is defined as ~\citep[see
e.g.][]{landau-fields,harko-lobo-book,franklin17,nuastase19}:

\begin{equation}
  T_{\alpha\beta} =\pm \frac{ 2 }{ \sqrt{-g} }\frac{ 
    \delta \left( \sqrt{-g} \, \lmatt 
    \right) }{ \delta g^{\alpha\beta}},
\label{energy-momentum}
\end{equation}

\noindent or, equivalently~\citep[see e.g.][]{mendoza2021matter}:

\begin{equation}
 T_{\alpha\beta} = \pm 2 \frac{ \delta \lmatt }{ \delta g^{\alpha\beta} } \mp
   g_{\alpha\beta} \lmatt,
\label{em-co}
\end{equation}

\noindent and in  contravariant representation:
\begin{equation}
 T^{\alpha\beta} = \mp 2 \frac{ \delta \lmatt }{ 
     \delta g_{\alpha\beta} } \mp g^{\alpha\beta} \lmatt.
\label{em-contra}
\end{equation}

\section{Non-Perfect Fluids}
\label{Fluids}

  The first law of thermodynamics for relativistic fluids can be written
as \citep[see e.g.][]{landau-fluids}:

\begin{equation}
  \mathrm{d}\left( \frac{ e }{ \rho } \right) =T \mathrm{d}\left( \frac{
  \sigma }{ \rho } \right) - p \mathrm{d}\left( \frac{ 1 }{ \rho } \right),
\label{first-law}
\end{equation}

\noindent where $\sigma=\rho s$ is the entropy density of the fluid with
$s$ the entropy per unit mass and $\rho$ the fluid or gas (baryonic) mass
density. The total energy density \( e = \rho c^2 + \xi \) contains the
rest energy density \( \rho c^2 \) and a pure internal energy density
term \( \xi \). The pressure and temperature are represented by $p$
and $T$, respectively.

 In the Eckart frame~\citep{eckart1948thermodynamics} for relativistic
fluids, the energy-momentum tensor of a non-perfect fluid is written as
\citep[see e.g.][]{rezzolla2013relativistic}:

\begin{equation}
  T^{\alpha \beta}= T^{\alpha \beta}_\text{pf}+ T^{\alpha
    \beta}_\text{npf},
\label{em0}
\end{equation}

\noindent where $T^{\alpha \beta}_\text{pf}$ and $ T^{\alpha
\beta}_\text{npf}$ refer to the perfect and non-perfect fluid
contribution, respectively. Equation~\eqref{em0} can be written
explicitly as \citep[see e.g.][]{rezzolla2013relativistic}:

\begin{equation}
  T^{\alpha \beta}=eu^\alpha u^\beta + (p+\Pi)h^{\alpha \beta} +q^\alpha
    u^\beta + q^\beta u^\alpha +\pi^{\alpha \beta},
\label{em-np}
\end{equation}

\noindent where $\Pi, q^\alpha$ and $\pi^{\alpha \beta}$ are, the
viscous bulk pressure, the heat flux and the anisotropic stress
tensor respectively. In equation \eqref{em-np} the projection tensor
is represented by $h^{\alpha \beta}=u^{\alpha} u^{\beta}-g^{\alpha
\beta}$.  The covariant divergence of the complete energy-momentum
tensor~\eqref{em0} is null, i.e.:

\begin{equation}
  \nabla_\alpha T^{\alpha\beta} = 0.
\label{null-divergence}
\end{equation}

\section{Matter Lagrangian for a Non-Perfect Fluid}
\label{matter-lagrangian}

In the Eckart frame the mass density current \( \rho u^\alpha \)
is parallel to the four velocity and in the continuity
equation takes the same form as for a perfect fluid, i.e.~\citep[see
e.g.][]{rezzolla2013relativistic}:

\begin{equation}
  \nabla_\alpha \left( \rho u^\alpha \right)=0,
\label{continuity}
\end{equation}

\noindent For this reason the variation of the mass density $\rho$
satisfies~\citep[see e.g.][]{mendoza2021matter}:

\begin{equation}
  \frac{ \delta \rho }{
    \delta g^{\alpha\beta} } =
    \frac{ 1 }{ 2 } \rho \left( g_{\alpha\beta} - u_\alpha u_\beta \right),
    \label{vr}
\end{equation}

\noindent for a non-perfect fluid in the Eckart frame.

In order to obtain the variation of the entropy per unit mass $s$
in terms of the variation of the metric tensor $g^{\alpha \beta}$ let us
multiply equation~\eqref{null-divergence} by \( u_\beta \) so that the
following energy equation is obtained:

\begin{gather}
    u_{\beta}\nabla_\alpha T^{\alpha \beta} =\nabla_\alpha ((e+p) u^\alpha ) - u^{\alpha}\nabla_\alpha p + \Pi \nabla_\alpha u^\alpha \nonumber \\+ 
    \nabla_\alpha q^\alpha + u^\alpha u_\beta \nabla_\alpha q^\beta +u_\beta \nabla_\alpha \pi^{\alpha \beta}.
    \label{energy}
\end{gather}
    
   Since the relativistic enthalpy \( \omega \) is defined as:

\begin{equation}
\omega := e+p=\rho (\mu +T s), 
\label{enthalpy}
\end{equation}

\noindent where $\mu := \omega/\rho -T s$ is the relativistic chemical 
potential, it follows from the first law of
thermodynamics~\eqref{first-law} that:

\begin{equation}
  \mathrm{d}\mu =\frac1\rho \mathrm{d}p - s \mathrm{d}T.
\label{first-law-mu}
\end{equation}

\noindent Using this, we can rewrite the energy equation~\eqref{energy} as:

\begin{equation}
  \begin{split}
    u^\alpha \nabla_\alpha s = &-\frac{\Pi}{T\rho}\nabla_\alpha u^\alpha
      -\frac{1}{T\rho}\nabla_\alpha q^\alpha -\frac{1}{T\rho}u_\beta
      u^\alpha \nabla_\alpha q^\beta 
      		\\
    &-\frac{1}{T\rho}u_\beta \nabla_\alpha \pi^{\alpha \beta}
  \end{split}
\label{t01}
\end{equation}

  Let us now assume  that the specific entropy $s$, the four velocity
$u^{\alpha}$ and thermodynamic fluxes $q^\alpha$ and $\pi^{\alpha \beta}$
depend only on the metric tensor $ g^{\alpha\beta}$ and not on its
derivatives $ \partial g^{\alpha\beta} / \partial x^\lambda $, so that:

\begin{equation}
  \frac{\partial}{\partial x^\mu}=\frac{\partial g^{\alpha
    \beta}}{\partial x^\mu}\frac{\partial}{\partial g^{\alpha \beta}}.
\label{op}
\end{equation}

Using this operator~\eqref{op} and the following relations: 

\begin{equation}
  u_\mu \frac{\partial u^\mu}{\partial g^{\alpha \beta}} = \frac12
    u_\alpha u_\beta,
\end{equation}

\begin{equation}
    \frac{\partial \sqrt{-g}}{\partial g^{\alpha \beta}} = -\frac12 \sqrt{-g} g_{\alpha \beta},
\end{equation}

\begin{equation}
    \frac{\partial g^{\mu \nu}}{\partial g^{\alpha \beta}} = \frac12 \left(\delta^\mu_\alpha \delta^\nu _\beta +\delta^\mu_\beta \delta^\nu _\alpha \right),
\end{equation} 

\begin{equation}
    \frac{\partial g_{\mu \nu}}{\partial g^{\alpha \beta}} = -\frac12 \left(g_{\mu \alpha} g_{\nu \beta}+g_{\mu \beta} g_{\nu \alpha}\right),
\end{equation}

\begin{equation}
    u_\mu \frac{\partial q^\mu}{\partial g^{\alpha \beta}} = \frac14 \left( q_\alpha u_\beta +q_\beta u_\alpha \right),
\end{equation}

\noindent in equation~\eqref{t01} it follows that:

\begin{equation}
 \begin{split}
  \frac{ \delta s }{ \delta g^{\alpha\beta} } = \frac{ \partial s }{
    \partial g^{\alpha\beta} }  &= \frac{ 1 }{ 2 } \frac{\Pi}{\rho T}\left(
    g_{\alpha\beta} - u_\alpha u_\beta \right)
		\\
   &- \frac{ 1 }{ 2 } \frac{1}{\rho T}\left( q_\alpha u_\beta + u_\alpha
    q_\beta +\pi_{\alpha \beta}\right),
  \end{split}
\label{vs}
\end{equation}

\noindent for a non-perfect fluid.  

  Let us now choose the specific entropy \( s \) and the mass density \(
\rho \) as the dependent variables for any given thermodynamical quantity,
so that:

\begin{equation}
  \frac{\delta }{\delta g^{\alpha \beta}} = \frac{\delta \rho}{\delta
  g^{\alpha \beta}}\eval{\frac{\partial }{\partial \rho}}_s  +
  \frac{\delta s}{\delta g^{\alpha \beta}} \eval{\frac{\partial }{\partial
  s}}_\rho.
\label{vl}
\end{equation}

  Substitution of the operator~\eqref{vl} into equation~\eqref{em-co}
and using relations~\eqref{vr} and~\eqref{vs}, it follows that:

\begin{equation}
 \begin{split}
  &T_{\alpha\beta} = \pm \left\{ \rho \eval{\frac{\partial \lmatt}{\partial
   \rho}}_s 
   + \frac{\Pi}{\rho T}\eval{\frac{\partial \lmatt}{\partial s}}_\rho
   \right\} \left( g_{\alpha\beta} - u_\alpha u_\beta \right) 
  		\\
  &\mp \frac{1}{\rho T}\eval{\frac{\partial \lmatt}{\partial s}}_\rho\left(
   q_\alpha u_\beta + u_\alpha q_\beta +\pi_{\alpha \beta}\right) \mp
   g_{\alpha\beta} \lmatt,
 \end{split}
\label{em1}
\end{equation}

\noindent for a non-perfect fluid. Direct comparison of
equations~\eqref{em-np} and \eqref{em1} imply that:

\begin{equation}
 \begin{split}
   &eu_\alpha u_\beta + (p+\Pi)h_{\alpha \beta} +q_\alpha u_\beta +
   q_\beta u_\alpha +\pi_{\alpha \beta} =
    \\ 
    &\pm \left\{ \rho \eval{\frac{\partial \lmatt}{\partial \rho}}_s 
    + \frac{\Pi}{\rho T}\eval{\frac{\partial \lmatt}{\partial s}}_\rho
    \right\} \left( g_{\alpha\beta} - u_\alpha u_\beta \right)
     \\ 
    &\mp \frac{1}{\rho T}\eval{\frac{\partial \lmatt}{\partial
    s}}_\rho\left( q_\alpha u_\beta + u_\alpha q_\beta +\pi_{\alpha
    \beta}\right)
    \mp g_{\alpha\beta} \lmatt.
 \end{split}
\label{diff-equation}  
\end{equation}

  The previous equation~\eqref{diff-equation} can be solved for \(
\lmatt \) in two different ways:

\begin{enumerate}[(i)]
  \item By multiplication of equation~\eqref{diff-equation} by \(
        u^\alpha u^\beta \) to obtain:
	\begin{equation}
	  \lmatt = \mp e.
	\label{e1}
	\end{equation}
  \item In order to obtain a system of equations for $\lmatt$, let us 
        first multiply equation~\eqref{diff-equation} by the metric tensor \(
        g^{\alpha\beta} \) to obtain:
	\begin{equation}
	  e - 3\left(p + \Pi\right)= \pm 3 \rho \eval{\frac{\partial
	  \lmatt}{\partial \rho}}_s \pm 3 \frac{\Pi}{\rho T}
	  \eval{\frac{\partial \lmatt}{\partial s}}_\rho \mp 4 \lmatt.
        \label{e2}
	\end{equation}
        and multiply equation~\eqref{diff-equation} by the projection tensor
        \( h^{\alpha}{ }_\lambda := u^\alpha u_\lambda - 
	\delta^\alpha{ }_\lambda \) to obtain:

\begin{equation}
  \begin{split}
    &- (p+\Pi)(u_\lambda u_\beta -g_{\lambda \beta} ) -(q_\lambda u_\beta +\pi_{\lambda \beta}) = 
		\\
    &\mp  \left(\lmatt- \rho \eval{\frac{\partial \lmatt}{\partial \rho}}_s -\frac{\Pi}{\rho T}\eval{\frac{\partial \lmatt}{\partial s}}_\rho\right)(u_\lambda u_\beta -g_{\lambda \beta} )
    \\
    &\pm \frac{1}{\rho T}\eval{\frac{\partial \lmatt}{\partial
    s}}_\rho\left( q_\lambda u_\beta  +\pi_{\lambda \beta}\right),
  \end{split}
\label{e3.0}
\end{equation}

Equating terms with \( (u_\lambda u_\beta -g_{\lambda \beta)}\)
and the ones with \( (q_\lambda u_\beta +\pi_{\lambda \beta})
\) in equation~\eqref{e3.0}, the following two relations are
obtained:

\begin{gather}
  p+\Pi = \pm  \left(\lmatt- \rho \eval{\frac{\partial \lmatt}{\partial
    \rho}}_s -\frac{\Pi}{\rho T}\eval{\frac{\partial \lmatt}{\partial
    s}}_\rho\right),
			\label{e3} \\
\intertext{and}
    \frac{1}{\rho T}\eval{\frac{\partial \lmatt}{\partial s}}_\rho =\mp 1.
			\label{e4}
\end{gather}
 
\end{enumerate}

The solution to the system of equations~\eqref{e2},~\eqref{e3} and~\eqref{e4} 
recovers the same value of \( \lmatt = \mp e  \) given in relation~\eqref{e1}.

With the value of \( \lmatt = \mp e \) the set of
equations~\eqref{e3}-\eqref{e4} can be written as:

\begin{equation}
    \eval{\frac{\partial e}{\partial \rho}}_s= \frac{e+p}{\rho},
\label{t1}
\end{equation}

\noindent and

\begin{equation}
  \eval{\frac{\partial e}{\partial s}}_\rho = \rho T,
\label{t2}
\end{equation}

\noindent which are consistent with the first law of 
thermodynamics~\eqref{first-law}.

\section{Final remarks}
\label{remarks}

  In this work we have shown that the value of the matter Lagrangian for a non-perfect fluid is 
\begin{equation}
  \lmatt = \mp e,
\label{matter-lagrangian-final}
\end{equation}

\noindent where the sing depends on the definition of the matter action and on the signature of the metric\footnote{For the choice of a
(\(-,+,+,+\)) signature, equation~\eqref{vr} turns into 
\citep[see e.g.][]{mendoza2021matter}:

\begin{displaymath}
  \frac{\delta \rho}{\delta g^{\alpha\beta}} = \frac{ 1 }{ 2 } \rho \left( g_{\alpha\beta} + u_\alpha
    u_\beta \right) ,
\end{displaymath}

\noindent and equation \eqref{vs} is now:

\begin{equation}
  \frac{ \delta s }{
    \delta g^{\alpha\beta} } =
    \frac{ 1 }{ 2 } \frac{\Pi}{\rho T}\left( g_{\alpha\beta} + u_\alpha
    u_\beta \right) + \frac{ 1 }{ 2 } \frac{1}{\rho T}\left( q_\alpha
    u_\beta + u_\alpha q_\beta +\pi_{\alpha \beta}\right).
\end{equation}

\noindent With all these results it follows that:

\begin{equation}
  \lmatt = \pm e.
\end{equation}
}
\footnote{In the Landau frame the energy momentum tensor \( T_{\alpha \beta}\)
and the particle flux density \( \rho_\alpha \) of a non-perfect fluid are 
respectively given by~\citep[see e.g.][]{landau-fluids}:
\begin{equation}
  T_{\alpha \beta}=(e+p)u_\alpha u_\beta -p g_{\alpha \beta} + 
    \tau_{\alpha \beta},
\end{equation}

\begin{equation}
\rho_\alpha = \rho u_\alpha +\nu_\alpha,
\end{equation}

\noindent where $\tau_{\alpha \beta}$ is the viscosity tensor and 
$\nu_\alpha$ is the thermal conduction of the fluid.

Analogously to the development shown in section \ref{matter-lagrangian},
the variation of the mass density $\rho$ and the entropy per unit mass
$s$ can be written as:

\begin{equation}
\frac{\delta \rho}{\delta g^{\alpha \beta}}=\frac{1}{2} \rho (g_{\alpha \beta} -  u_\alpha u_\beta )- \frac{1}{4 } (\nu_\alpha u_\beta + u_\alpha \nu_\beta ),
\label{lfvr}
\end{equation}

\noindent and

\begin{equation}
\frac{\delta s}{\delta g^{\alpha \beta}}=\frac{1}{4 \rho} \left(s + \frac{\mu}{T}\right) (\nu_\alpha u_\beta + u_\alpha \nu_\beta)- \frac{1}{2 \rho T} \tau_{\alpha \beta},
\label{lfvs}
\end{equation}

\noindent respectively. With all this, the same result of
equation \eqref{matter-lagrangian-final} is also obtained in the
Landau frame.}. This value is consistent with the first law of
thermodynamics~\eqref{first-law}.

Note that the same result for the matter Lagrangian is obtained for an
ideal fluid~\citep[see e.g.][]{mendoza2021matter}. The difference resides
in the value of the total energy density for each fluid, which is the
addition between the rest mass energy density plus the pure internal
energy density.

\section*{Acknowledgements} 
\label{acknowledgements} 
This work was supported by a PAPIIT DGAPA-UNAM grant IN110522. SM and
SS acknowledge support from CONAHCyT (26344,1228643).

\bibliographystyle{apsrev4-2}
\bibliography{mendoza-silva}

\end{document}